# The Prevailing Role of Hotspots in Plasmon-Enhanced Sum-Frequency Generation Spectroscopy

Laetitia Dalstein[1,2], Christophe Humbert[1], Maroua Ben Haddada[3], Souhir Boujday[3], Grégory Barbillon[4], Bertrand Busson[1,*]

1) Laboratoire de Chimie Physique, CNRS, Univ. Paris-Sud, Université Paris-Saclay, Bâtiment 201 P2, F-91405 Orsay, France

2) Institute of Physics, Academia Sinica, Taipei 11529, Taiwan

3) Sorbonne Université, CNRS, Laboratoire de Réactivité de Surface (LRS), 4 place Jussieu, F-75005 Paris, France

4) EPF-Ecole d'Ingénieurs, 3 bis rue Lakanal, F-92330 Sceaux, France

*Corresponding author: bertrand.busson@u-psud.fr

ABSTRACT

The plasmonic amplification of non-linear vibrational sum frequency spectroscopy (SFG) at the surfaces of gold nanoparticles is systematically investigated by tuning the incident visible wavelength. The SFG spectra of dodecanethiol-coated gold nanoparticles chemically deposited on silicon are recorded for twenty visible wavelengths. The vibrational intensities of thiol methyl stretches extracted from the experimental measurements vary with the visible color of the SFG process and show amplification by coupling to plasmonics. Since the enhancement is maximal in the orange-red region rather than in the green, as expected from the dipolar model for surface plasmon resonances, it is attributed mostly to hotspots created in particle multimers, in spite of their low surface densities.

TOC

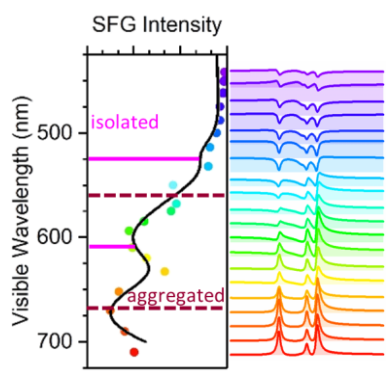

**TEXT**

Surface-Enhanced Raman Spectroscopy (SERS),[1] benefitting from local electromagnetic enhancements in the visible through plasmonics, is a widely used optical tool giving access to chemical footprints of molecules at surfaces. With the same assets as SERS as for chemical sensitivity and wavelengths ranges, infrared-visible Sum-Frequency Generation (SFG) spectroscopy is in addition intrinsically surface specific as a consequence of its symmetry properties. It has indeed proved its ability to extract static,[2,3] dynamic,[4,5] structural,[6,7] and orientational information[8–10] through vibrational excitations from molecules adsorbed on planar surfaces, as well as from a wide range of systems like air-water,[6,11–16] solid-liquid or buried interfaces.[11,17–25] For nanostructured materials, it should also become an useful tool to monitor the chemistry taking place at the surfaces of nano-objects,[26] essential for the comprehensive design of nanosensors[27] and nanocatalysts.[28]

Despite the apparent discrepancy between the spherical symmetry of nanoparticles and SFG selection rules, the vibrational SFG response of molecules decorating nanospheres grafted on a substrate has been calculated and shows enhancements in the visible range (as in SERS) due to the local excitation of surface plasmon resonances.[29] This is also true for the molecules constituting the grafting monolayer sandwiched between the particles and the substrate.[29] Amplification of SFG spectroscopic signals has been shown on long deposited gold cylinders,[30] for which the inversion symmetry is greatly broken. In the case of spheres, no direct evidence of such enhancement has been provided and, for gold particles, the amplification factors are expected to be rather small. Consequently, SERS amplification has never been measured, to our knowledge, for isolated small gold spheres, which makes the challenge even more interesting. Several authors have shown the possibility to monitor by SFG molecules at the surfaces of plasmonic spherical particles,[31–40] and to make the difference between thiols adsorbed on the particles and silanes used to graft them on silicon surfaces.[35] Showing vibrational enhancement in the small (i.e. for diameters up to a few tenths of nanometers) particle case will prove that SFG is a promising tool for a local spectroscopic analysis of all nanostructured materials, with a high sensitivity to species located in the very local field around the nanostructures.

In this letter, we provide experimental evidence of a plasmonic enhancement of SFG vibrational signatures of terminal methyl moieties of dodecanethiol (DDT) grafted on gold nanoparticles. The evolution of the vibrational intensities as a function of the visible wavelength shows a direct effect of the optical properties of the particles. It does not primarily relate to the surface plasmon resonances of the isolated particles but, despite their low surface density, to the plasmonic properties of the hotspots between nearby particles.

The SFG intensity radiated by a surface is given by[41,42]

$$I(\omega_{SFG} = \omega_{vis} + \omega_{IR}) = \frac{8\pi^3 (\omega_{SFG})^2}{c^3 \cos^2 \theta_{SFG}} \left| \chi^{(2)}_{eff} \right|^2 I(\omega_{vis}) I(\omega_{IR}) \qquad (1)$$

where $\theta_{SFG}$ is the emission angle of the SFG reflected beam and $I(\omega)$ stands for the beam intensity at frequency $\omega$ (refractive index of air is unity). The SFG signal experimentally measured is proportional to its total radiated intensity and, after taking into account the wavelength-dependent efficiency of the detection setup,[43] provides the dispersion of $|\chi^{(2)}_{eff}|^2$, $\chi^{(2)}_{eff}$ being the effective, Fresnel-corrected, second order non-linear susceptibility.[41] The complex third rank order tensor $\chi^{(2)}_{eff}$ sums up one Lorentzian term per molecular vibration ($\chi^{(2)}_{eff,mol}$) and one contribution ($\chi^{(2)}_{eff,NR}$) from the inorganic components, essentially the silicon substrate, labelled non-resonant (NR) with respect to the IR wavelength but varying in amplitude and phase with the visible wavelength. Interferences between these two contributions arise because of the summation of the $\chi^{(2)}_{eff}$ amplitudes, whereas only the total intensity is experimentally measured:

$$I(\omega_{SFG}) \propto \left| \chi^{(2)}_{eff,mol} + \chi^{(2)}_{eff,NR} \right|^2 \quad (2)$$

with

$$\chi^{(2)}_{eff,mol} = \sum_{vibrations} \frac{A_i}{\omega_{IR} - \omega_i + i\Gamma_i} \quad (3)$$

For the analysis of the coupling between plasmonics and non-linear optics, in line with our previous study on DDT adsorbed on gold,[42,44,45] twenty vibrational spectra were recorded on DDT-coated gold nanoparticles deposited on silicon while tuning the visible color between 442 and 710 nm (see Experimental Section). Gold particles (diamater 13.5 nm) were grafted on silicon through an alkoxysilane layer (APTES) modified with mercaptoundecanoic acid (MUA). Such particles have a cuboctahedral shape,[46] and thiols form small self-assembled layers on the facets.[47,48] As on planar gold, long alkanethiol chains favor lateral interactions and lead to long range molecular order.[48] We have checked by SFG that DDT led to a better organization than shorter alkanethiols (see Supplementary Information for sample preparation and characterizations). The corresponding $|\chi^{(2)}_{eff}|^2$ are shown in Figure 1. As is clearly seen in the red visible spectral range, they exhibit three vibrations corresponding to the stretching modes of DDT methyl endgroups at 2888 cm$^{-1}$ (symmetric stretch, *ss*), 2951 cm$^{-1}$ (Fermi resonance, *FR*) and 2974 cm$^{-1}$ (antisymmetric stretch, *as*). It is known that incomplete silanization reaction of alkoxysilanes on silicon may lead to poisoning from residual $CH_2$ and $CH_3$ modes from the grafting layer.[35,49,50] The presence of $CH_2$ vibration modes is therefore possible, their amplitudes showing the degree of disorder inside both the grafting and functionalization layers. We have checked that the sample was mostly free from these effects (see SI for details).

The spectra can be classified into three regions as a function of the visible wavelength: in the orange-red zone, high intensity CH$_3$ modes distinctly appear above a very small non-resonant background. At the blue side of the visible spectrum, the resonant amplitudes decrease whereas the non-resonant grows, as the SFG photons may excite electronic transitions inside silicon, producing spectra where Lorentzian-shaped peaks strongly interfere with a high level of background. In the intermediate zone (green region), the amplitudes of the resonant and non-resonant signals are comparable, resulting in pronounced interference with profiles evolving from peaks to dips through derivative shapes.

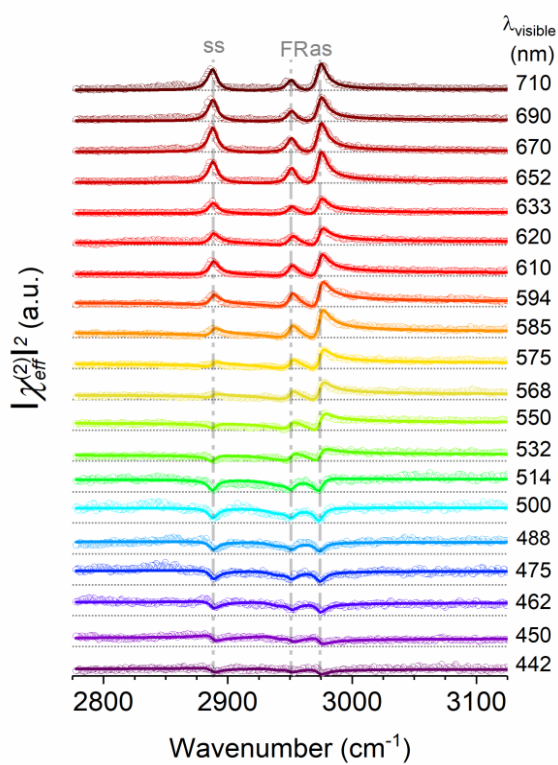

*Figure 1. Series of SFG spectra of the Si/AuNP/DDT interface for twenty visible wavelengths (open circles). Lines are fits according to equations 2 and 3 (see text for details). All spectra share a common scale.*

Determining the effects of plasmonics on the molecular vibrational amplitudes $A_i$ requires extracting them from the experimental spectra. On planar gold,[42,45] the vibration amplitudes do not depend on the visible color and can be used as an internal reference for the gold signals in amplitude and in phase. The goal here is oppositely to measure their variations when the visible color is tuned, in order to quantify the influence of plasmon resonances on the molecular non-linear susceptibility. The presence of a non-resonant background, varying in amplitude and phase from blue to red, makes it necessary to fit the spectra according to Equations 2 and 3.[41,42] Consistently fitting twenty spectra with three resonances, sharing a common width $\Gamma$, and a non-resonant with unknown amplitude and phase has been shown difficult in the past.[42,51] Details are provided in the SI.

In Figure 2A, we show the evolutions of the vibrational intensities of the three $CH_3$ modes as a function of the visible wavelength. The coherent lineshapes for the three resonances show the reliability of the fits, as we expect similar evolutions for symmetric and antisymmetric stretches.[29] The intensities show a regular increase from blue to green, with the presence of two broad maxima around 580 and 660 nm. This fact proves that plasmonics influences indeed the molecular SFG response in the vicinity of the nanoparticles and produces a net enhancement of the SFG intensities.

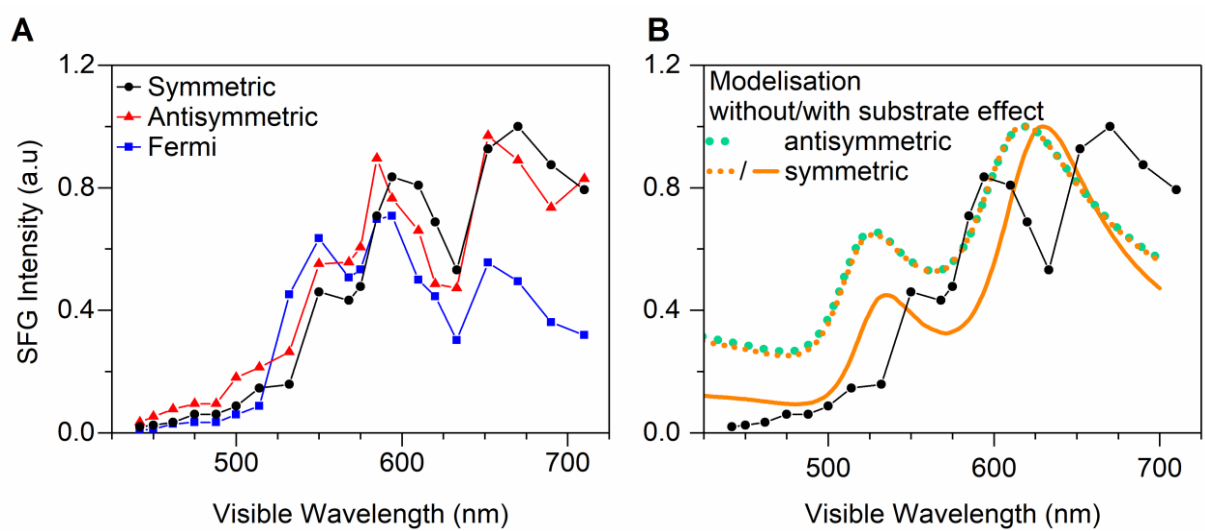

*Figure 2. Intensities of the methyl vibrations versus the visible wavelength (A) Experimental values from the spectra of Figure 1. (B) Simulations of the experimental dispersion for symmetric and antisymmetric stretches, with and without inclusion of the substrate effects ($\varepsilon_m=1.44$, $\alpha=35°$). Curves are scaled as indicated to ease comparison.*

The observed trend however differs from the predictions of a straight plasmonic model for gold nanospheres, for which plasmon excitation by the visible beam lies around 510nm (see Figure S1D for the visible absorption spectrum of the sample), and 620 nm for the SFG beam. We compare the experimental results with the predictions of the dipolar model developed in a previous paper[29] and summarized in the SI. In the calculations, we have set all the parameters to their experimental values (surface density of particles and particle diameter, see SI for details). The DDT layer thickness and $CH_3$ hyperpolarizability properties have been discussed before.[29] The free parameters of the model are essentially the refractive index $n_m$ of the DDT layer (dielectric function $\varepsilon_m = n_m^2$) around the particles, whose increase leads to a redshift of the plasmon peaks, and the half-aperture of the cone ($\alpha$) describing the surface of the spheres covered with DDT, as thiol functionalization is performed after grafting the particles on the surface (Figure S5). We recall here that a spherical nanoparticle fully covered with molecules produces no measurable signal, only the symmetry breaking induced by the coverage cone making SFG production possible. Including the influence of the silicon substrate also induces an additional redshift. In Figure 2B, we compare the experimental intensities for the symmetric stretch to the simulations with the reference values of $n_m = 1.2$ and $\alpha = 35°$ (theoretical limit value for DDT in contact with the grafting layer[29], Figure S5), with and without influence of the substrate. The comparison illustrates that these models do not account for the experimental results as they do not reproduce the positions of the peaks, their relative intensities and the decrease towards the blue side. This discrepancy may arise from the incorrect account of the influence of the silicon substrate on the electric fields, which is stronger below the particle than above it, creating an up-down asymmetry. We also considered hybrid particles, for which this asymmetry is recovered (see SI for details), but could not reproduce the experimental behavior either (Figure S6B). Finally, taking into account the distribution of sphere radii and local environments (through $\varepsilon_m$) only marginally modifies the calculated dispersion and does not change the conclusions.

The obvious experimental fact is that the experimental enhancement is redshifted with respect to the plasmonic model in the dipolar approximation for isolated spheres. Conversely, it is well-known that longitudinal plasmonic coupling between two (or more) nanoparticles creates hotspots, characterized by a redshift of the plasmon resonance[52] and a great increase of the local electric fields between the particles.[53] Depending on the conditions, in particular the distance between the particles, amplification of two orders of magnitude may be expected at the longitudinal plasmon wavelength for interparticle distances below one nanometer.[53–55] As both the visible and SFG wavelengths may lie in its vicinity, we may expect enhancements up to four magnitude orders as compared to isolated spheres. In addition, the distribution of the electric field in a nanoparticle dimer strongly breaks the spherical symmetry of the ensemble, thus contributing proportionally more to SFG production than an isolated sphere. Finally, we note that the positions of the experimental SFG maxima are coherent with the plasmonic response of multimers as measured by absorbance spectroscopy (~600nm) when many aggregates are present at the surfaces, as illustrated on Figure S2. Experimental evidences thus suggest that hotspots created by particle multimers greatly contribute to SFG production, even if their surface density is low (but not vanishing). Considering their expected high electric field enhancement factors, those may compensate the small number of aggregates on the surface (which accounts for their absence in absorbance measurements).

It is in principle possible to model the electric field distribution around a particle multimer using Mie theory.[56] However, this goes beyond the dipolar approximation used so far and, more important, it requires adjusting unknown parameters to reproduce the experimental conditions, namely the multimer geometries (interparticle separation, which may fluctuate a lot with great consequences on plasmonics,[57,58] number of aggregated particles and relative positions) and their surface densities (the small number of aggregates would require a big amount of large microscopy images for a statistically relevant experimental measurement [27]). Here, we propose to account for the experimental behavior in a phenomenological way. We take advantage of the similarity of plasmonic properties between longitudinally coupled spheres and isolated spheres surrounded by a high dielectric function host (Figure 3A). In both cases, the surface plasmon resonance experiences a redshift, while keeping a reproducible shape as a function of the wavelength of light. Of course, the origins of both redshifts are different, but as far as electronic resonance is concerned, they share a common behavior. In a qualitative approach, we take into account the existence of dimers and multimers by adding to the dipolar contribution of isolated spheres a second contribution described by a strongly redshifted dipolar distribution with high values of the dielectric function $\varepsilon_m$.

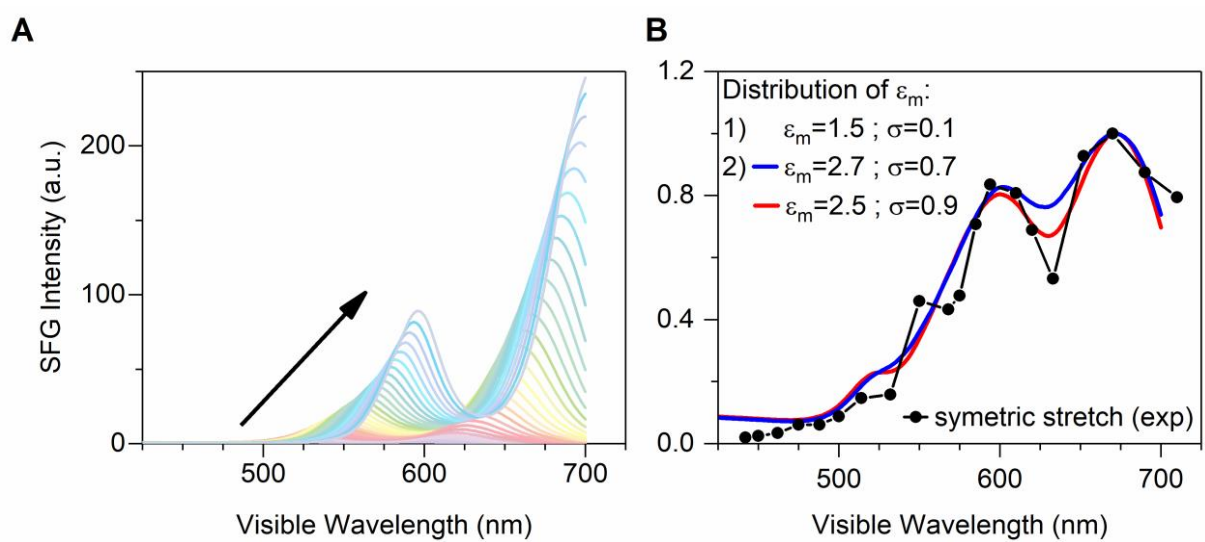

*Figure 3. Evolution of the SFG response versus the visible wavelength (A) As a function of $\varepsilon_m$. The graph displays the SFG intensities for thirty values of $\varepsilon_m$, varying from 1.0 to 4.0 along the arrow.(B) Simulations of the experimental dispersion for symmetric methyl stretch using two distributions in $\varepsilon_m$ as described in the text.*

We therefore optimized the fit of the experimental intensities using a sum of two Gaussian distributions of hybrid spheres: a narrow one centered at low values of $\varepsilon_m$ represents the isolated spheres surrounded by the DDT layer, whereas a wider one centered at higher values of $\varepsilon_m$ accounts for the diversity of couplings in the multimers due to the varieties of geometries and interparticle distances. The best fits are shown on Figure 3B, with the first Gaussian centered at $\varepsilon_m = 1.5$ with a width (standard deviation) of 0.1, the second one at either 2.5 (width 0.9) or 2.7 (width 0.7), and $\alpha = 30°$, close to the limit value. These values are consistent with the usual refractive index for organic monolayers as determined by SFG spectroscopy ($n_m \approx 1.2$) and with the peak positions for the redshifted resonances. The overall plasmonic amplification over the visible range is recovered in this way. Its shape comes from an overlap between plasmonic resonances at the visible and SFG wavelengths, both from isolated and interacting particles, the latter being broadened by a large distribution of interaction geometries, mostly interparticle distances.

SFG spectroscopy of molecules adsorbed at the surfaces of grafted nanoparticles is strongly influenced by plasmonics. By probing the visible spectral range through the measurement of a series of vibrational SFG spectra, we have evidenced that the intensities of the molecular vibrational resonances depend on the visible wavelength. The molecular SFG signal may be resonantly amplified, even in the unfavorable case of small gold spherical nanoparticles for which the optical enhancement remains weak. The plasmonic enhancement cannot be accounted for by isolated gold particles alone. Despite the low surface density of multimers and aggregates, the observed redshift proves that these predominantly contribute to plasmonic coupling in the SFG process, as a consequence of the high local electric fields and symmetry breaking associated to hotspots between nearby particles. This hotspot effect is well-known for other enhanced spectroscopies, but is shown here for the first time in the case of sum-frequency generation spectroscopy, extending the field of non-linear plasmonics to molecular vibrational spectroscopy.

EXPERIMENTAL SECTION

The two-colour SFG set-up is analogous to the one described before.[42,59] Briefly, a ~10-ps vanadate laser source is used to independently generate, after temporal shaping and amplification, tunable IR and visible beams through two dedicated optical parametric oscillators (OPO) based on $LiNBO_3$ and BBO nonlinear crystals, respectively. The presence of a tunable visible beam (in the range 440-710 nm) makes it possible to measure the SFG response in the visible spectral range. SFG signals are produced in a reflection geometry at the surface of the sample for *ppp*-polarization combination, with incidence angles of 55° and 65° for the visible and IR beams, respectively. SFG photons are filtered from reflected and scattered visible in a double grating monochromator and measured with a photomultiplier. At each step, vibrational SFG spectra are recorded as a function of the IR wavenumber in the 2800-3100 $cm^{-1}$ region for a fixed visible color. In this spectral range, CH stretching vibration modes are detected and, for symmetry reasons, $CH_3$ modes from the methyl endgroups of alkanethiols are the dominant source of vibration resonances.

**Associated Content**

**Supporting Information**

Sample preparation and characterization, details on the curve fitting, summary of the theoretical model, influence of the substrate on the plasmonic properties.

**Author Information**

**Corresponding author**

*E-mail: bertrand.busson@u-psud.fr

**ORCID**

Laetitia Dalstein: 0000-0002-2591-5421

Christophe Humbert: 0000-0001-9173-3899

Souhir Boujday: 0000-0002-5500-0951

Grégory Barbillon : 0000-0003-0009-7343

Bertrand Busson: 0000-0002-7652-1772

**Author contributions**

L.D. and B.B. designed the research project. L.D. and C.H. performed the SFG experiments. L.D. performed the UV-Visible measurements, M. BH. and S.B. prepared the samples and performed the TEM characterizations, G.B. performed the SEM characterization. All the authors discussed the results and wrote the manuscript.

**Notes**

The authors declare that they have no competing interest.

**Acknowledgments**

We are very grateful to Dr Jean-Paul Hugonin and Prof. Jean-Jacques Greffet (Laboratoire Charles Fabry, Institut d'Optique Graduate School, Palaiseau, France) for the T-matrix calculations.

Supplementary Information for

# The Prevailing Role of Hotspots in Plasmon-Enhanced Sum-Frequency Generation Spectroscopy

Laetitia Dalstein[1,2], Christophe Humbert[1], Maroua Ben Haddada[3], Souhir Boujday[3], Grégory Barbillon[4], Bertrand Busson[1,*]

1) Laboratoire de Chimie Physique, CNRS, Univ. Paris-Sud, Université Paris-Saclay, Bâtiment 201 P2, F-91405 Orsay, France

2) Institute of Physics, Academia Sinica, Taipei 11529, Taiwan

3) Sorbonne Université, CNRS, Laboratoire de Réactivité de Surface (LRS), 4 place Jussieu, F-75005 Paris, France

4) EPF-Ecole d'Ingénieurs, 3 bis rue Lakanal, F-92330 Sceaux, France

*Corresponding author: bertrand.busson@u-psud.fr

1. **Sample preparation and characterization**
2. **Fitting procedure**
3. **Theory of the SFG response of a thiol decorated particle in the dipolar model**
4. **Influence of the substrate on the plasmonic properties**

## 1. Sample preparation and characterization

### 1.1. Sample structure

We aimed at producing nanostructured interfaces with controlled chemical and optical properties, also showing long-term stability. On the other hand, experimental procedures should remain accessible and aim towards routine production. We therefore opted for the chemical route, with a calibrated synthesis of particles, deposited afterwards on a silicon surface, chosen in place of glass because of its high reflectivity. Silicon was chemically modified by silane grafting. The particles were in a last step functionalized with probe thiol molecules. The two key steps concern the grafting of particles on the silane layer, and the functionalization by thiol molecules. First, the dipolar interpretation of SFG amplification in terms of plasmonics relates to non-interacting plasmons, that is from isolated particles. The idea here was not to maximize the net amplification but rather to show that SFG still provides quantitative chemical information on the molecules even in the most unfavorable case of small, isolated gold nanoparticles. We thus wanted to get rid of aggregation as much as possible, while still maintaining a high surface density of spheres at the surface of silicon. Second, thiol grafting at the surface should be as efficient as possible, in order to ensure a high surface density of molecular markers on the spheres, thus a highly ordered thiol overlayer. For both criteria, stability in time and during the various steps of the study (e.g. thiol functionalization, SFG spectroscopy) was also required.

### 1.2. Sample preparation

The following criteria should be met by the synthetized nanostructured interfaces: small spherical gold particles with a narrow dispersion in size, deposited on silicon with a high surface density but the lowest density of aggregates, efficiently functionalized by thiols in order to form an ordered molecular layer at their surfaces with a long-term stability.

All chemicals, except when mentioned, were provided by Sigma-Aldrich and used as is. An aqueous solution of gold nanoparticles was prepared according to the classical citrate method modified with tannic acid.[1,2] The particles were rather monodisperse as expected, and TEM images show an average diameter of 13.5 ± 1 nm (Figures S1A and S1B).

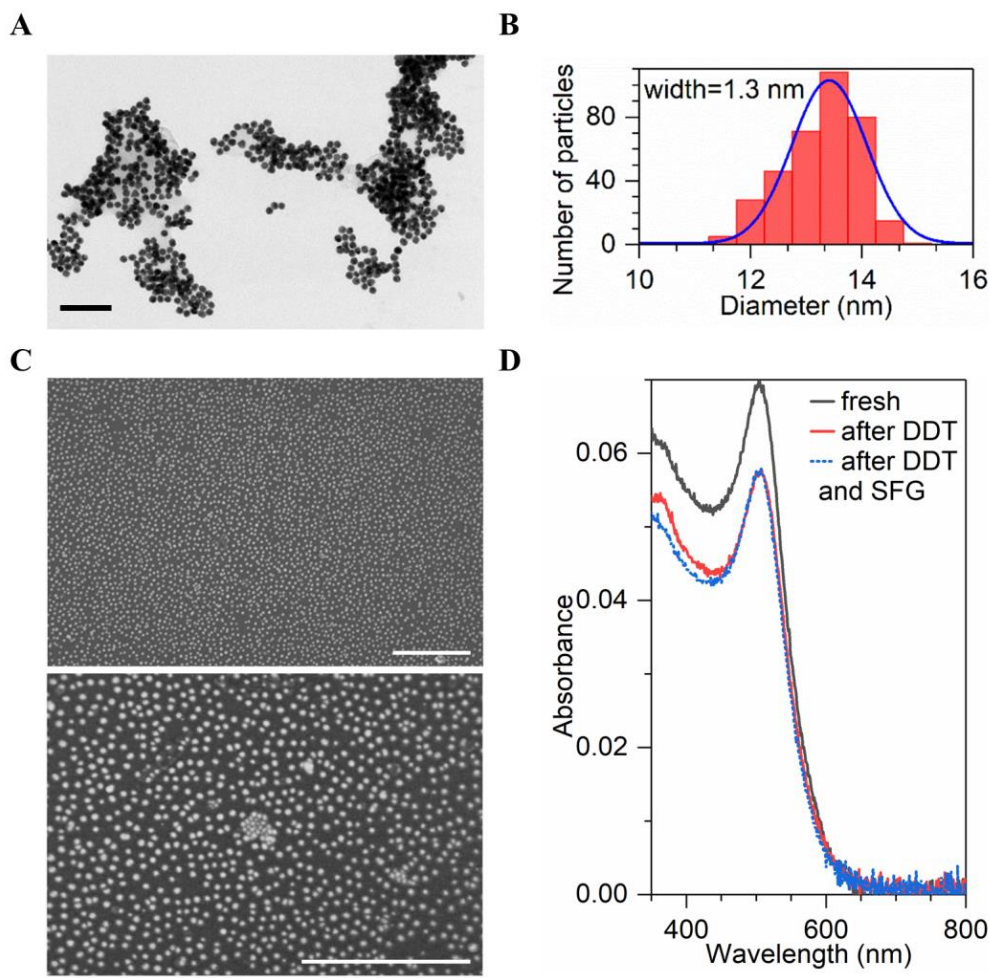

*Figure S1: A. TEM image of gold nanoparticles deposited on a copper grid (Scale bar = 100 nm). B. Histogram of nanoparticle diameters as measured by TEM. C. SEM images (Scale bar = 500 nm) and D. Absorbance spectra after grafting of gold nanoparticles on APTES-MUA modified silicon wafer (sample 2).*

Silicon substrates (Sil'tronix wafers, (100) oriented, *n*-doped, resistivity 20 to 30 mΩ.cm) were first washed with acetone then ethanol (from Analar Normapur) in ultrasonic bath, then cleaned in a piranha solution (sulfuric acid and hydrogen peroxide from Carlo Erba) and finally treated with UV-ozone. The surfaces were first silanized with (3-aminopropyl) triethoxysilane (APTES) in anhydrous toluene (Analar Normapur) at high temperature.[3] In a second step, grafted APTES was reacted with an ethanol solution of 11-mercaptoundecanoic acid (MUA) activated by NHS-EDC protocol.[2] After ethanol rinsing, gold nanoparticles were grafted on the APTES-MUA functionalized wafers by dipping into the nanoparticle solution for 30 minutes. It has been shown[2] that the MUA overlayer combined to the short immersion time led to the best compromise between the surface densities of isolated and aggregated particles. We realized several Si/AuNP samples and used UV-visible spectroscopy to selected one with both a high surface coverage and low aggregation rate (sample 2, see Figures S1D and S2). Finally, grafted particles were functionalized by dipping in a $10^{-2}$ M thiol solution (propanethiol, heptanethiol, octanethiol, dodecanethiol (DDT) or octadecanethiol (ODT)) in ethanol for 12 hours. After thiol functionalization, scanning electron microscopy images confirm that, although dimers and small aggregates are present, their surface densities are low (Figure S1C). The surface density of particles for the selected Si/AuNP/DDT sample is measured as $8.33 \; 10^{10}$ particles.cm$^{-2}$.

### 1.3. Nanoparticle aggregation

It is essential to define quality criteria for the design of the samples. We have investigated the particle grafting step in our previous study,[2] using either APTES or APTES/MUA, and two immersion times in the nanoparticle solution, showing that APTES/MUA and 30 minutes are the best choice. Using these conditions, the surface density of particles and aggregates is still difficult to accurately control. Several samples were prepared and checked by both scanning electron microscopy and *in-situ* UV-visible spectroscopy in reflection geometry, the latter representing the standard tool to determine plasmonic properties on silicon. Spectra were recorded at an incidence angle of 55°, and several examples are provided in Figure S2. The main peak is produced by the excitation of the surface plasmon resonance of isolated spheres (around 510 nm). The second feature relates to the redshifted longitudinally coupled plasmons, appearing as a dip close to 600 nm as a consequence of refractive index contrast with silicon. The amplitude of the main peak relates to the surface density of grafted nanoparticles, whereas the second one measures the aggregation density. We note that the absence of the second feature warrants that the surface density of oligomers remains below a given threshold, although it is difficult to get rid of all aggregates over a macroscopic sample. We have chosen the sample 2 for this study, which gives the highest amplitude for the surface plasmon resonance of isolated particles, without any measurable contribution from the aggregates.

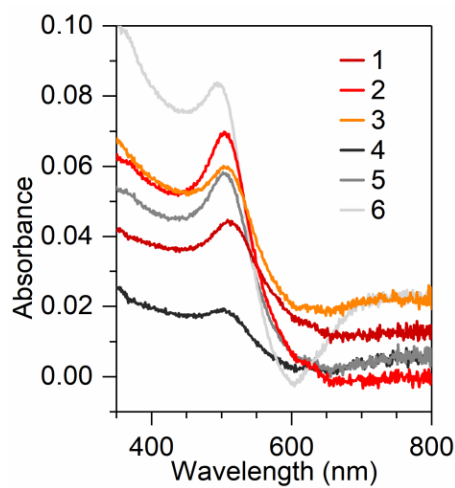

*Figure S2: Examples of absorbance spectra of nanoparticle grafted silicon surfaces using 30 min (1,2,3) and 60 min (4,5,6) dipping times.*

### 1.4. Chemical functionalization

Plasmonic amplification will be experienced by any molecule grafted on the particles' surfaces, but it requires a high level of molecular order to be accurately quantified. In order to select the best molecule for that purpose, we used SFG spectroscopy in order to characterize the efficiency of molecular functionalization (Figure S3A). We compared five alkanethiols with increasing chain lengths. Short alkanethiol chains (i.e. propane, heptane and octanethiols) gave rise to unreliable coating. Most of the times, SFG signals from their $CH_3$ end groups were not distinguishable enough from the signals produced by the silanes in the same wavenumber range, and included $CH_2$ contributions showing some disorder in the molecular layers. In addition, the $CH_3$ vibration widths account for a mixture of methyl populations at the interface. On the other hand, long octadecanethiol chains disturbed the particle grafting and resulted in poor quality interfaces. DDT was by far the most efficient, reliable and reproducible thiol chain for this study. The selected sample 2 was efficiently chemically functionalized since, after DDT grafting, the three $CH_3$ modes may be clearly evidenced and overwhelm the silane SFG features (Figure S3B). We note that SFG spectroscopy proves a reliable tool for estimating molecular order at the surface of nanoparticles.

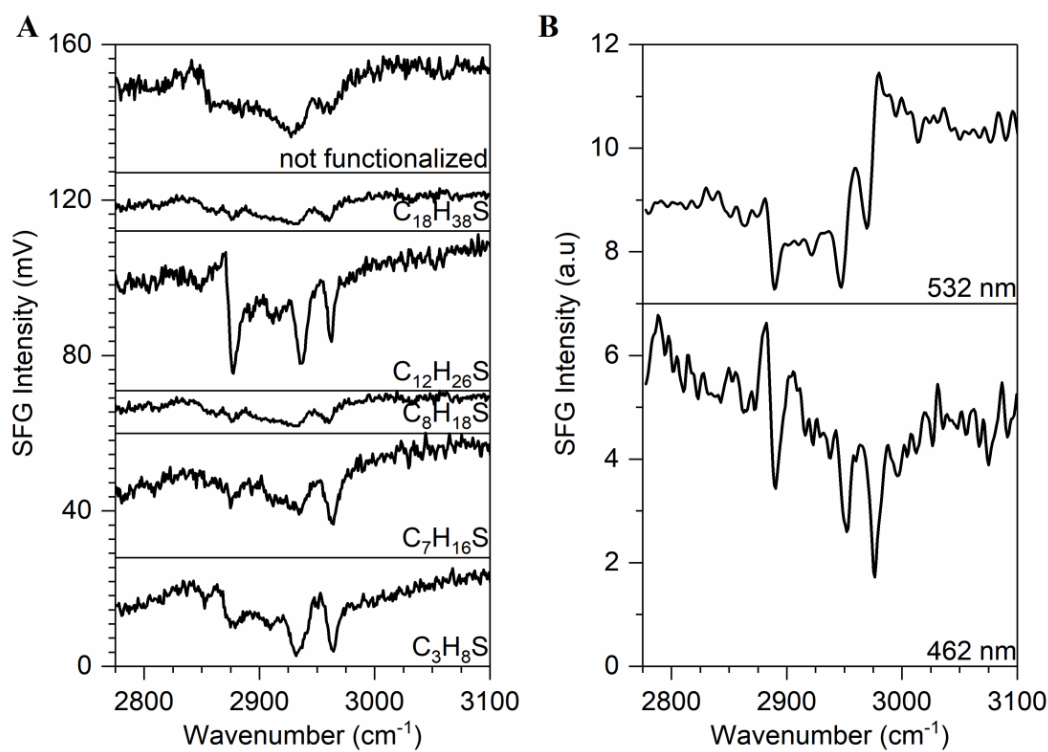

*Figure S3: A. Representative SFG spectra of thiol-functionalized gold nanoparticles on silicon for five alkanethiols. B. Two SFG spectra of the DDT-functionalized sample 2.*

### 1.5. Stability

Since the samples are submitted to intense laser irradiation in the visible and the infrared for long periods of time, it is crucial to check their long-term stability both under beam and by natural ageing. For this work, we encountered a variety of situations as far as ageing was concerned. Some samples were very robust, even standing for hours under the intense beam of the CLIO free electron laser,[4] whereas some others rapidly experienced disordering in air.[2] An example of a slow increase of molecular disorder under beam is shown below in Figure S4: after recording several SFG spectra, the layer starts disordering and we clearly see a distinct feature of $CH_2$ symmetric stretch appearing, and the $CH_2$ antisymmetric mode starts interfering with the $CH_3$ Fermi resonance.

This phenomenon did not happen for Sample 2. We assessed its stability during the recording of the series of SFG spectra by regularly checking the 532 nm SFG spectrum, and ensuring that no parasitic $CH_2$ signal appeared (Figure S3B). In addition, the UV-visible reflectivity spectra recorded before and after the whole SFG session showed no evidence of any damage related to plasmonic properties (Figure S1D). In conclusion, this sample has met all the required criteria for nanoparticle deposition, thiol functionalization and chemical ageing.

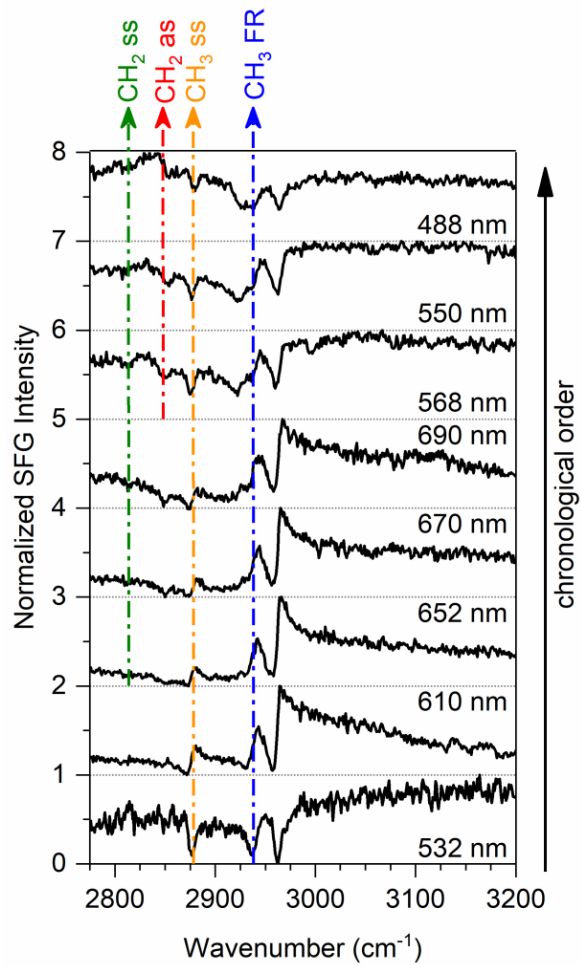

*Figure S4*: Series of SFG spectra of a DDT-functionalized interface, showing a regular increase in molecular disorder (green and red arrows). Time increases from bottom to top. The sample was submitted to laser irradiation for several hours.

## 2. Fitting procedure

The fits have first been performed using three $CH_3$ vibration modes. In a second step, we have seen that it was necessary to take into account a small contribution from a fourth peak at 2939 cm$^{-1}$, characteristic of antisymmetric $CH_2$ stretches in order to ensure the regularity of all fit parameters over the whole visible range. It most probably arises from remnants of the silanization step in the APTES layer[2] or from a slightly disordered MUA contribution (see also the unfunctionalized spectrum in Figure S3a).

In addition, the fitting procedure becomes problematic in the green region, where resonant and non-resonant compare in amplitude. The symmetric stretch almost disappears from the spectra between 532 and 575 nm and fitting parameters become very sensitive. Fitting for these spectra was improved when an additional resonance with a small amplitude was added around 2850 cm$^{-1}$, standing for the $CH_2$ symmetric stretch, as it changes the $\chi^{(2)}$ phase in the vicinity of the 2888 cm$^{-1}$ peak. However, we have checked that fitting without this fifth resonance did not introduce significant differences for the other parameters. All the fits shown in Figure 1 and amplitudes displayed in Figure 2 therefore follow from a fitting procedure with four resonances.

### 3. Theory of the SFG response of a thiol decorated particle in the dipolar model

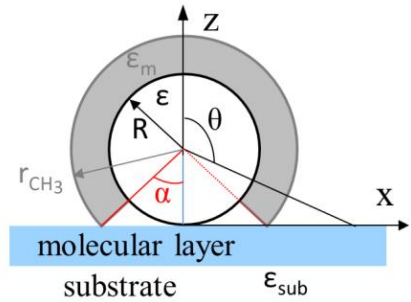

*Figure S5: scheme of one particle (radius R, dielectric function ε) decorated with DDT molecules (grey, thickness $r_{CH3}$ – R, dielectric function $ε_m$) deposited on a substrate (dielectric function $ε_{sub}$) through a grafting layer (blue). Angle α quantifies the extent of molecular coverage on the particle.*

We summarize here the principles of the theoretical model presented in Ref 5. For molecules at the surface of a sphere, the relationships between the molecular local field $\mathbf{E}^{local}$ and the applied far field $\mathbf{E}^{far}$ are summarized in Scheme 1. The far field is modified by the reflectivity of the substrate to give rise to the interface field $\mathbf{E}^0$. Both are conveniently described by their Cartesian coordinates in the laboratory frame (**x**,**y**,**z**) related to *p* and *s* polarizations of light. $\mathbf{E}^0$ is then expressed in the spherical coordinates ($\mathbf{u}_\theta,\mathbf{u}_\varphi,\mathbf{u}_r$). The local field effects, which include here plasmonic amplification, are taken into account in the spherical frame through a matrix $\tilde{\mathbf{\Lambda}}$ (defined in Ref. 5), to obtain the local field $\mathbf{E}^{local}$ eventually experienced by the molecule. This field is more conveniently expressed in the molecular frame (**a**,**b**,**c**).

$$\mathbf{E}^{local,(a,b,c)} \xrightarrow{\mathbf{D}_{mol}} \mathbf{E}^{local,(\theta,\varphi,r)} \xrightarrow{\tilde{\Lambda}} \mathbf{E}^{0,(\theta,\varphi,r)} \xrightarrow{\mathbf{D}^0} \mathbf{E}^{0,(x,y,z)} \xrightarrow{F_{ijk}} \mathbf{E}^{far,(x,y,z)}$$

$$\beta_{\alpha\beta\gamma} \xrightarrow{\mathbf{D}_{mol}+\diamond_{mol}} \beta_{\mu\nu\xi} \xrightarrow{\tilde{\Lambda}} \beta^0_{\mu\nu\xi} \xrightarrow{\mathbf{D}^0+\diamond_{\theta,\varphi}} \chi^{(2)}_{ijk}, \beta^{sphere}_{ijk} \xrightarrow{F_{ijk}} \chi^{(2)}_{eff}$$

*Scheme 1: Definitions and relationships between electric fields, first hyperpolarizabilities and second-order nonlinear susceptibilities involved in the calculation of the SFG response of molecules at the surface of a sphere.*

The actual calculation of all quantities involved in Scheme 1 is performed from the microscopic ($\mathbf{E}^{local}$) to the macroscopic ($\mathbf{E}^{far}$) levels. The molecular hyperpolarizabilities $\beta_{\alpha\beta\gamma}$ are first averaged on molecular Euler angles at each point of the surface of the sphere. The average molecular hyperpolarizabilities $\beta_{\mu\nu\xi}$ experience the plasmonic enhancement, then are summed up to produce a molecular hyperpolarizability for the whole sphere, weighted by their surface density and corrected by Fresnel factors to produce the final effective susceptibility. The formulas required at each stage may be found in the original reference 5.

### 4. Influence of the substrate on the plasmonic properties

The model presented in the previous paragraph may be extended to take into account the presence of the substrate.[5] It creates an asymmetry between SFG responses parallel and perpendicular to the substrate. Nevertheless, the dipolar model introduces no asymmetry between the lower and upper sides of the particles. However, a calculation of the electric field amplitude around the particle using a recursive T-matrix algorithm for a simplified stratified system[8] (Figure S6A) shows indeed that the presence of the substrate enhances the electric fields more intensely under the particle, which introduces a second symmetry breaking in the model. We have taken this effect into account in a simple way, by separating the particle in two halves, one below and one above the center, for which the SFG amplitudes were calculated with and without influence of the substrate, respectively, then summed up. When α is tuned, the lower contribution varies whereas the upper one remains constant. The results for these "hybrid" particles still differ from the experimental results (Figure S6B). Contrary to the symmetric situation, maximal intensity is obtained for a fully covered particle as a result of an overall increase of the number of emitters. For increasing angles α, the amplitudes reach a minimum for which compensation between upper and lower halves is maximum, then grow again as the upper half starts to dominate the response at higher angles.

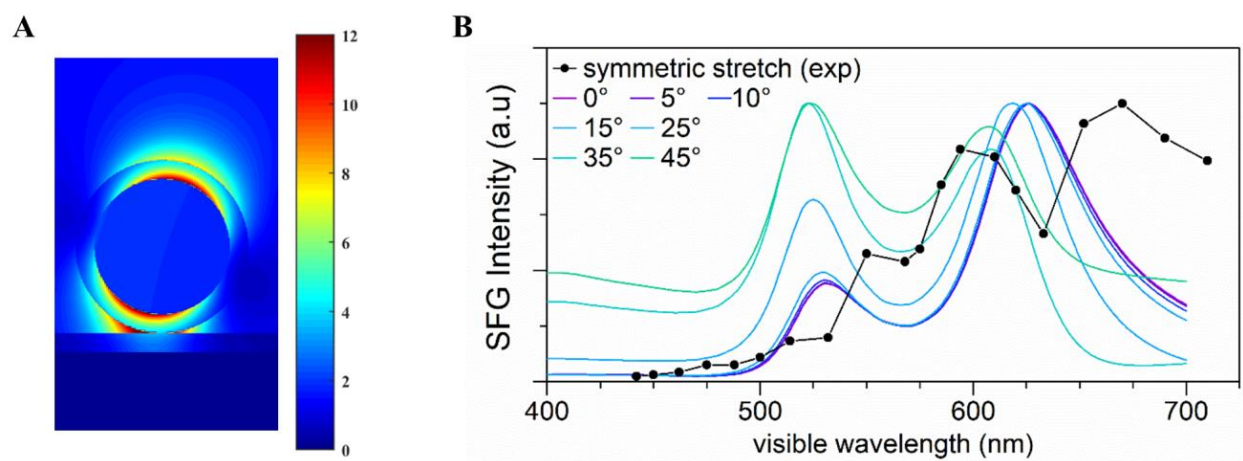

Figure S6: A. Normalized electric field intensity $|E/E_0|^2$ at a wavelength of 500 nm. B. Evolution of the SFG intensities for hybrid particles as a function of α varying from 0° to 45° ($\varepsilon_m = 1.44$).